\pdfoutput=1
\documentclass[aps,prd,nofootinbib,twocolumn,citeautoscript,showkeys]{revtex4-1} 
\usepackage{amsmath}   
\usepackage{graphicx}  % needed for figures
\usepackage{dcolumn}   % needed for some tables
\usepackage{bm}        % for math
\usepackage{amssymb}   % for math
\usepackage{epsfig}
\usepackage{epstopdf}  
\usepackage{subfigure}
\usepackage[utf8]{inputenc}
\usepackage{mdframed}
\usepackage{amsmath}
\usepackage[utf8]{inputenc}
\usepackage{cancel}
\usepackage[normalem]{ulem}
\usepackage{wasysym}
\usepackage[hidelinks,bookmarks=false]{hyperref} 
\hypersetup{pdfstartview=FitH,pdfhighlight=/O}
\usepackage{tabu}
\usepackage[toc]{appendix}
\usepackage{color,soul} 
\usepackage{datetime}

\newcommand{\beq}{\begin{equation}}
\newcommand{\eeq}{\end{equation}}
\newcommand{\bal}{\begin{aligned}}
\newcommand{\eal}{\end{aligned}}
\newcommand{\rmd}{\mathrm{d}}

\begin{document}

\title{Corrections to Extremal Black Holes from Iyer-Wald Formalism}
\author{Lars Aalsma} 
\email{laalsma@wisc.edu}
\affiliation{Department of Physics, University of Wisconsin-Madison, 1150 University Avenue, Madison, Wisconsin
53706, USA}
%\keywords{}
%\pacs{}

\begin{abstract}
We present a general method of computing corrections to the extremality bound and entropy of Kerr-Newman black holes due to an arbitrary perturbation using the Iyer-Wald formalism. In this method, corrections to the extremality bound are given by an integral over an effective stress tensor which, in particular cases of interest, reduces to the usual stress tensor. This clarifies the relation between extremality corrections and energy conditions. In particular, we show that a necessary condition to decrease the mass of an extremal black hole in a canonical ensemble, as required by the weak gravity conjecture, is that the perturbation violates the dominant energy condition. As an application of our method, we compute higher-derivative corrections to charged black holes in anti-de Sitter space and Kerr black holes.
\end{abstract}

\maketitle

\section{Introduction}
In the absence of experiments that directly probe Planckian energies, a regime where quantum gravity effects become important, it is useful to understand if quantum gravity puts any constraints on low-energy effective field theories (EFTs). While one might naïvely expect that any such effect is suppressed by the Planck scale and is therefore irrelevant at low energies, it turns out that the structure of quantum gravity is more interesting than that. Even in systems with low energies and curvatures, quantum gravity still seems to put some constrains on EFTs. Two famous examples of this effect are the black hole information paradox, which suggests corrections to the standard semiclassical picture of black holes to be consistent with unitarity, and the absence of global symmetries in quantum gravity.

The goal of the swampland program is to make precise what constraints quantum gravity puts on EFTs (see \cite{Brennan:2017rbf,Palti:2019pca} for reviews). One such constraint is the weak gravity conjecture (WGC), which states that any EFT with a U(1) gauge field coupled to gravity should contain a state that has a charge-to-mass ratio that exceeds the one for extremal black holes in that theory \cite{Arkani-Hamed:2006emk}. Kinematically, this allows extremal black holes to shed their charge by emitting superextremal states. The so-called mild form of the WGC suggests that black holes themselves can be the states that satisfy the WGC, which is possible because higher-derivative corrections modify the black hole extremality bound \cite{Kats:2006xp}. Depending on the sign of the Wilson coefficients multiplying the higher-derivative corrections, this allows extremal black holes to have a charge-to-mass ratio that exceeds the one in the uncorrected theory, thereby satisfying the WGC. This mild form of the WGC has been studied for many different black hole solutions \cite{Cheung:2018cwt,Hamada:2018dde,Cheung:2019cwi,Loges:2019jzs,Bellazzini:2019xts,Cano:2019oma,Cano:2019ycn,Charles:2019qqt,Jones:2019nev,Cremonini:2019wdk,Cremonini:2020smy,Loges:2020trf,Melo:2020amq,Aalsma:2020duv,Arkani-Hamed:2021ajd,Cremonini:2021upd} and it has been demonstrated that, at least under some assumptions about the UV theory from which the higher-derivative corrections originate, unitarity and causality constrain the signs to be compatible with the WGC \cite{Hamada:2018dde,Bellazzini:2019xts,Arkani-Hamed:2021ajd}. Additionally, in particular cases the mild form and particle form of the WGC can be shown to be intimately connected \cite{Aalsma:2019ryi}.

In \cite{Cheung:2018cwt,Hamada:2018dde}, an interesting perspective has been put on the mild form of the WGC by demonstrating that corrections to the extremality bound in a canonical ensemble are related to corrections to the entropy in a microcanonical ensemble. This entropy/extremality relationship follows quite generally from thermodynamics \cite{Loges:2019jzs,Goon:2019faz} and reformulates the WGC as a statement that (higher-derivative) corrections should increase the microcanonical entropy of an extremal black hole. 

In this paper, we will further investigate this relation for four-dimensional Kerr-Newman black holes and give a new derivation of it based on the Iyer-Wald formalism. Similar to a thermodynamic approach, this has the advantage that it is not necessary to explicitly solve for the corrected metric which can be a laborious task. Using this approach we obtain a relation that relates corrections to the extremality bound to an integral over an effective stress tensor whose positivity properties determines the sign of the corrections. We show that this integral is directly related to corrections to the area of the black hole that, in many cases of interest, gives the dominant contribution to the microcanonical black hole entropy. The effective stress tensor captures any perturbation to the action and not just higher-derivative corrections. In special cases, the effective stress tensor is equivalent to the standard stress tensor. This occurs for Reissner-Nordström and Kerr black holes and we show that in those cases a necessary condition to decrease the mass of an extremal black hole in the canonical ensemble is that the perturbation to the stress tensor violates the dominant energy condition (DEC). However, in general this condition is not sufficient.

As an application and consistency check of our method, we compute higher-derivative corrections to four-dimensional charged black holes in anti-de Sitter space and Kerr black holes. In those cases, we investigate to what extent energy conditions can be used to constrain the sign of the higher-derivative corrections. We find that a violation of the DEC only fixes the sign of the $(F_{ab}F^{ab})^2$ operator. This is the same operator constrained by unitarity and combining this with the null energy condition (NEC) the WGC follows for charged black holes that are small with respect to the AdS radius. However, due to the angular dependence of rotating black holes, energy conditions do not constrain the sign of higher-derivative corrections to Kerr black holes.

This paper is organized as follows. In Sec. \ref{sec:IyerWald} we use the Iyer-Wald formalism to derive a general relation between corrections to Kerr-Newman black holes and the effective stress tensor. Continuing, in Sec. \ref{sec:ExtremalCorrections} we discuss how this relation relates to energy conditions and apply it to four-dimensional higher-derivative corrected charged black holes in AdS and Kerr black holes. Finally, we comment on the formulation of the WGC in AdS space in Sec. \ref{sec:AdSWGC} and conclude in Sec. \ref{sec:conclusion}.

\section{Black Hole Corrections from Iyer-Wald} \label{sec:IyerWald}

\subsection{Iyer-Wald Formalism}
In this section, we use the Iyer-Wald formalism \cite{Wald:1993nt,Iyer:1994ys} to derive how an arbitrary perturbation corrects four-dimensional Kerr-Newman black holes. We start by considering a general Lagrangian $d$-form $L(\phi)$ that depends on an arbitrary set of matter fields $\phi$. Varying with respect to the matter fields yields the following general form.
\beq
\delta L(\phi) = E(\phi) + \rmd \theta(\phi,\delta\phi) ~.
\eeq
$E(\phi)$ collectively denotes the equations of motion and $\rmd \theta(\phi,\delta\phi)$ are boundary terms. When the equations of motion are satisfied $E(\phi)=0$. We define a symplectic current $(d-1)$-form as follows:
\beq
\omega(\phi;\delta_1\phi,\delta_2\phi) = \delta_1\theta(\phi,\delta_2\phi) - \delta_2\theta(\phi,\delta_1\phi) ~.
\eeq
For some vector field $\xi$ we can define a Noether current $J_\xi(\phi)$:
\beq \label{eq:Noether1}
J_\xi(\phi) = \theta(\phi;{\cal L}_\xi\phi) - \iota_\xi L(\phi) ~.
\eeq
Here $\iota_\xi$ denotes the interior product. At the same time, it can be shown \cite{IyerWald1995} that the Noether current can also be written as
\beq \label{eq:Noether2}
J_\xi(\phi) = C_\xi  + \rmd Q_\xi ~.
\eeq
Here $C_\xi \equiv \xi^aC_a$ captures the constraints of the theory such that $C_a=0$ when $E(\phi)=0$. $ Q_\xi$ is the Noether charge associated with $\xi$. By varying the Noether current and equating \eqref{eq:Noether1} and \eqref{eq:Noether2} we obtain
\beq
\bal \label{eq:mainrelation}
\rmd\left(\delta Q_\xi-\iota_\xi\theta(\phi,\delta\phi)\right) &= \omega(\phi;\delta\phi,{\cal L}_\xi\phi) - \delta C_\xi \\
&-\iota_\xi(E(\phi)\delta\phi) ~.
\eal
\eeq
By assuming that $\xi$ is a symmetry of the action, such that ${\cal L}_\xi\phi=0$, we find that when integrated over a Cauchy surface $\Sigma$ the relation \eqref{eq:mainrelation} can be written on shell as
\beq \label{eq:finalgeneralrelation}
\int_{\partial\Sigma}\left(\delta Q_\xi - \iota_\xi\theta(\phi,\delta\phi)\right) = -\int_\Sigma \delta C_\xi ~. 
\eeq
We will now evaluate this relation for the case of primary interest, which is Einstein-Maxwell theory.

\subsection{Einstein-Maxwell Theory}
The Lagrangian of  Einstein-Maxwell theory is
\beq
L = \frac1{2}\left(R-2\Lambda\right)\epsilon - \frac14 F_{ab}F^{ab}\epsilon ~.
\eeq
Note that we work in units where $8\pi G_d = 1$ and $\epsilon$ is the volume $d$-form. By explicitly varying the action we find the equations of motion
\beq
\bal
0&=G_{ab}+\Lambda g_{ab} + T^A_{ab} ~,\\
0&= \nabla_aF^{ab} ~,
\eal
\eeq
and the symplectic potential of the metric and gauge field
\beq
\bal
\theta^g_{abc} &= \frac{\epsilon_{abcd}}{2}g^{de}\left(\nabla^f\delta g_{ef}-g^{hi}\nabla_e\delta g_{hi}\right) ~,\\
\theta^A_{abc} &= -\epsilon_{abcd} F^{de}\delta A_e ~.
\eal
\eeq
The electromagnetic stress tensor is
\beq
T^{A}_{ab} = F_{ac}F_b^{\,\,c} - \frac14g_{ab}F_{cd}F^{cd} ~.
\eeq
Using \eqref{eq:Noether1} we find that the Noether current can be written precisely in the form $\eqref{eq:Noether2}$. The constraints are given by
\beq
\bal
C^g_{abc}  &= \epsilon_{abce}\xi^d(T^g)_d^{\,\,e} ~,\\
C^A_{abc} &= \epsilon_{abce} (\xi^dA_d)j^e ~,
\eal
\eeq
where the tensor $T^g_{ab}$ is defined by subtracting the electromagnetic part from the stress tensor and we introduced the electromagnetic current $j^a$:
\beq
\bal
T^g_{ab} &= G_{ab}+\Lambda g_{ab}-T^A_{ab} ~,\\
j^a &= \nabla_bF^{ab} ~.
\eal
\eeq
The Noether charges are given by
\beq
\bal
Q_{ab}^g &= \frac{\epsilon_{abcd}}{2}\nabla^c\xi^d ~,\\
Q_{ab}^A &= - \frac{\epsilon_{abcd}}{2} F^{cd}A_e\xi^e ~.
\eal
\eeq
Let us consider a stationary solution to the source-free equations of motion (that is $T^g_{ab}=j_a=0$). We assume the existence of a Killing vector $K^a = t^a + \Omega \phi^a $ with $t^a$ a timelike Killing vector, $\phi^a$ the axial Killing vector and $\Omega$ the angular potential. Notably this is the case for Kerr-Newman black holes, whose metric is given by
\beq \label{eq:KerrNewman}
\bal 
\rmd s^2 &= -\frac{\Delta(r)}{\rho(r)^2}\left(\rmd t-a\sin^2\theta\rmd\phi\right)^2 + \rho(r)^2\left(\frac{\rmd r^2}{\Delta(r)}+\rmd\theta^2\right) \\
&+\frac{\sin^2\theta}{\rho(r)^2}\left((r^2+a^2)\rmd\phi-a\rmd t\right)^2 ~.
\eal
\eeq
The gauge field is given by
\beq
A=-\frac{\sqrt{2}rr_Q}{\rho(r)^2}\left(\rmd t-a\sin^2\theta\rmd\phi\right) ~,
\eeq
and
\beq
\bal
\rho(r)^2 &= r^2 + a^2\cos^2\theta ~,\\
\Delta(r) &= r^2 - r_sr + a^2 + r_Q^2 ~,
\eal
\eeq
with $a=J/M$,  $r_s = M/(4\pi)$ and $r_Q = Q/(4\sqrt{2}\pi)$. The two horizons of the black hole, obtained by solving $\Delta(r)=0$, are given by
\beq
r_\pm =\frac12\left(r_s \pm \sqrt{r_s^2-4(a^2+r_Q^2)}\right) ~.
\eeq
The angular and electric potential are
\beq
\bal
\Omega &= \frac{a}{\left(\frac{M}{8\pi}\right)^2+a^2} ~, \\
\Phi &=\frac{2 M Q}{(8\pi a) ^2+M^2} ~.
\eal
\eeq
We can now evaluate \eqref{eq:finalgeneralrelation}. To do so, we note that $\partial\Sigma$ consists of two parts: a surface at infinity and the bifurcation surface of the black hole, which is the surface where the past and future horizon intersect. We will assume that the matter fields fall off sufficiently rapidly at infinity such that they do not contribute to the surface integral at infinity. This implies a particular gauge choice for the gauge potential $A_a$. Although this might lead to singular behaviour of $A_a$ on the horizon, its variation $\delta A_a$ is smooth \cite{Sorce:2017dst}, and this does not pose any further problems. Using the explicit form of the metric, see \eqref{eq:KerrNewman}, we find
\beq 
\int_\infty\left(\delta Q_K - \iota_K\theta(\phi,\delta\phi)\right) = \delta M - \Omega\delta J ~,
\eeq
where $M$ and $J$ are the ADM mass and angular momentum, respectively. Although \eqref{eq:KerrNewman} corresponds to an asymptotically flat black hole this relation also holds for AdS black holes. The matter fields will contribute to the integral over the bifurcation surface $B$. For the gravitational part this leads to the well-known result \cite{Wald:1993nt}
\beq
\int_B\left(\delta Q^g_K - \iota_K\theta^g\right) =\kappa\delta A_B ~.
\eeq
Here $\kappa$ is the surface gravity at the horizon and $\delta A_B$ is variation of the bifurcation surface.\footnote{Strictly speaking, only nonextremal black holes have bifurcation surfaces so to obtain a similar expression for extremal black holes we should take a limit from a nonextremal black hole. This is implicit in what follows.} To arrive at this relation we used $\left.K^a\right|_B=0$. Because the temperature of a black hole is given by $T=\kappa/(2\pi)$, we can write the right-hand side as $T\delta S_{\rm BH}$. It is important to note that this is the variation of the Bekenstein-Hawking entropy and not the Wald entropy, because possible higher-derivative corrections are considered to be perturbations that appear in $\delta C_K$. Similarly, for the electromagnetic part we find
\beq
\bal
\int_B\left(\delta Q^A_K - \iota_K\theta^A\right) &= \frac{\Phi}{2}\int_B\delta\left(\epsilon_{abcd}F^{cd}\right) ~, \\
&= \frac{\Phi}{2}\delta Q ~.
\eal
\eeq
Here $\Phi \equiv -\left.A_aK^a\right|_B$ is the electric potential at the horizon. To arrive at the right-hand side, which we recognize as being proportional to the electric flux $\delta Q$ through the horizon, we again used $\left.K^a\right|_B=0$. However, note that $\Phi$ is constant. Computing the variation of $C_K$ and performing an integration by parts \eqref{eq:finalgeneralrelation} becomes
\beq \label{eq:Iyer-Waldrelation}
\bal
&\delta M - \Omega\delta J - T\delta S_{\rm BH} - \Phi\delta Q = \\
&\qquad-\int_\Sigma\epsilon_{ebcd}\left[\delta (T^g)_a^{\,\,e}+F_{af}\delta F^{ef}\right]K^a ~.
\eal
\eeq
From this relation we can read off corrections to the different quantities of the black hole. Before we do so, let us make a remark. Although we did not necessarily assume that the Kerr-Newman black holes we are considering are only electrically charged, possible magnetic charges do not appear in \eqref{eq:Iyer-Waldrelation}. The reason for this, as was recently nicely explained in \cite{Mitsios:2021zrn}, is that the Iyer-Wald formalism only picks up the variation of charges that are associated with a gauge symmetry. This means that if we would like to consider magnetically charged or dilatonic black holes, we would need to take into account additional variations that lead to the conserved charges associated with the magnetic and scalar field. We leave this generalization to future work. Alternatively, one could use a thermodynamic approach based on the Euclidean action \cite{Loges:2019jzs} to compute such corrections.

\subsection{Corrections to Extremal Black Holes}
We will now see how \eqref{eq:Iyer-Waldrelation} can be employed to deduce corrections to the black hole extremality bound. First of all, we find it convenient to write the right-hand side as an ``effective’’ stress tensor
\beq
\delta T_{ab}^{\rm eff} \equiv\delta T^g_{ab}+F_{ac}\delta F_b^{\,\,c} ~.
\eeq
This quantity can be interpreted as the stress tensor of some perturbation in special cases that we will discuss later. We see this by rewriting the effective stress tensor as
\beq
\delta T_{ab}^{\rm eff} = \delta T_{ab} + \left(\frac12g_{ab}F_{cd}\delta F^{cd} - F_{ac}\delta F_b^{\,\,c}\right) ~,
\eeq
where the total stress tensor is the sum of the gravitational and electromagnetic part: $ \delta T_{ab} =  \delta T^g_{ab}+ \delta T^A_{ab}$. Only when the terms in brackets vanish, this effective stress tensor coincides with the standard stress tensor of the perturbation we are considering. To explicitly evaluate the integral over the stress tensor, it is useful to write it has
\beq
\int_\Sigma\epsilon_{ebcd}\delta \left(T_a^{\,\,e}\right)^{\rm eff}K^a = -\int_\Sigma\tilde\epsilon_{bcd}\delta T_{ae}^{\rm eff}K^an^e ~,
\eeq
where $\tilde\epsilon$ is the volume form on $\Sigma$ and $n^a$ the timelike unit normal to $\Sigma$. The relation of interest now takes the form
\beq \label{eq:finalrelation}
\delta M - \Omega\delta J - \Phi\delta Q - T\delta S_{\rm BH}= \int_\Sigma\tilde\epsilon_{bcd}\delta T_{ae}^{\rm eff}K^an^e ~.
\eeq
To evaluate \eqref{eq:finalrelation} we use the explicit form of the Kerr metric given in \eqref{eq:KerrNewman}. As mentioned before, we use the asymptotically flat form, but the same relation holds in AdS space. If we now define
\beq
\Xi:= M - \Omega J - \Phi Q ~,
\eeq
we note that extremal Kerr-Newman black holes without perturbation ($\delta T_{ab}^g=j_a=0$) obey
\beq \label{eq:extcond}
\delta\Xi = \delta M - \Omega\delta J - \Phi\delta Q = 0~.
\eeq
Thus, we see that when we send $T\to 0$ the correction to the integral over the effective stress tensor determines the correction to the extremality bound. Also, if we focus on black holes that obey $\delta \Xi = 0$, the correction to the entropy is given by minus the integral over the effective stress tensor. In summary, we have
\beq
\bal \label{eq:extremalcorrections}
\lim_{T\to0}\left(\delta M - \Omega\delta J - \Phi\delta Q\right) &=\lim_{T\to0} \int_\Sigma\tilde\epsilon_{bcd}\delta T_{ae}^{\rm eff}K^an^e  ~,\\
\left(T\delta S_{\rm BH}\right) &= -\int_\Sigma\tilde\epsilon_{bcd}\delta T_{ae}^{\rm eff}K^an^e~.
\eal
\eeq
The second relation involving the entropy seems similar to the relation between the microcanonical entropy and the correction to the Gibbs free energy (see for example Eq. (17) of \cite{Melo:2020amq}), but we stress again that \eqref{eq:extremalcorrections} involves the Bekenstein-Hawking entropy and not the Wald entropy. To relate our correction to the Wald entropy, one can use the first law $\delta M = T\delta S_{\rm Wald} + \Omega\delta J + \Phi\delta Q$, where $\delta S_{\rm Wald}$ is the variation of the Wald entropy, which leads to
\beq
T\left(\delta S_{\rm Wald} - \delta S_{\rm BH}\right) = \int_\Sigma\tilde\epsilon_{bcd}\delta T_{ae}^{\rm eff}K^an^e~.
\eeq
From the first law, it is clear that, when $\delta\Xi=0$, we obtain the derived correction to the Bekenstein-Hawking entropy.

While these equations already give a relationship between entropy and extremality corrections, we note that when $\delta\Xi=0$ the entropy correction must diverge in the limit $T\to0$ to yield a finite result. For this reason it is more insightful to evaluate the entropy corrections in a microcanonical ensemble (fixed mass, angular momentum and electric charge). From \eqref{eq:extremalcorrections} we then find that the microcanonical entropy is given by
\beq
\left(T\delta S_{\rm BH}\right)_{M, J, Q}  = -\int_\Sigma\tilde\epsilon_{bcd}\delta T_{ae}^{\rm eff}K^an^e ~.
\eeq
At the same time, the mass correction at $T=0$ in a canonical ensemble (fixed temperature, angular momentum and electric charge) is given by
\beq 
\lim_{T\to0}\left(\delta M\right)_{T,J,Q} = \lim_{T\to0}\int_\Sigma\tilde\epsilon_{bcd}\delta T_{ae}^{\rm eff}K^an^e ~. 
\eeq
Thus, we can relate the mass correction in the canonical ensemble in the extremal limit ($T\to 0$) to the microcanonical entropy correction in the extremal limit ($\Xi\to 0$) :
\beq \label{eq:massentropyrelation}
\lim_{T\to0}\left(\delta M\right)_{T,J,Q} = -\lim_{\Xi\to 0}\left(T\delta S_{\rm BH}\right)_{M,J,Q} ~.
\eeq
Similar relationships between entropy and extremality corrections have appeared before in the literature in various guises and have been derived by studying explicit corrections to black hole solutions as well as by using a thermodynamic approach \cite{Cheung:2018cwt,Hamada:2018dde,Cheung:2019cwi,Loges:2019jzs,Goon:2019faz,Cremonini:2019wdk,Aalsma:2020duv,Arkani-Hamed:2021ajd}

Let us now comment upon the validity and use of \eqref{eq:massentropyrelation}. First, it should be stressed that the order of limits in the entropy correction is important. We should first consider a variation of the entropy, for example with respect to a small expansion parameter proportional to the correction we are considering, and subsequently take the limit $\Xi\to 0$. As has been carefully explained in \cite{Cremonini:2019wdk} (see also \cite{Reall:2019sah}) this is important because these two limits do not necessarily commute: the microcanonical entropy need not to be analytic in the expansion parameter when $\Xi = 0$. Despite this subtlety, even when we first consider the limit $\Xi\to0$ and then perform the variation with a suitable expansion parameter, there still is a relationship between entropy corrections in a microcanonical ensemble and mass corrections in the canonical ensemble \cite{Cremonini:2019wdk} that is only modified by an order one constant. 

Second, one might be puzzled about topological terms, like the Gauss-Bonnet term in four dimensions. Because these terms appear as total derivatives when varying the action they do not contribute to the equations of motion, the effective stress tensor, or the mass and entropy correction. However, because they appear as boundary terms they have support on the bifurcation surface of the black hole and therefore do contribute to the Wald entropy. This is consistent with \eqref{eq:massentropyrelation} which only involves the Bekenstein-Hawking entropy, but seemingly in conflict with entropy/extremality relationships involving the Wald entropy \cite{Goon:2019faz}. The resolution lies in the fact that in a microcanonical ensemble, higher-derivative corrections only shift the temperature of a black hole when its area is corrected. Topological terms do not contribute to the area, so even when $\delta S$ is finite $\lim_{\Xi\to 0}(T\delta S)_{M,J,Q}=0$ because the temperature goes to zero in this limit.

This does imply, as remarked in \cite{Bobev:2021oku}, that there is no direct relationship between corrections to the extremality bound and the sign of the Wilson coefficients multiplying topological terms in the action. In particular, the WGC does not follow from requiring entropy corrections to be positive in the presence of topological terms. Nonetheless, we still believe \eqref{eq:massentropyrelation} to be interesting because in many cases of interest the correction to the temperature of an extremal black hole in a microcanonical ensemble scales as $T\sim {\cal O}(\sqrt{\alpha_i})$, where $\alpha_i$ denotes a Wilson coefficient. In that case, the shift in the area also scales with a square root leading to $\delta S_{\rm BH}\sim {\cal O}(\sqrt{\alpha_i})$ such that $\delta S = \delta S_{\rm BH} + {\cal O}(\alpha_i) $.  Possible corrections from topological terms appear at order ${\cal O}(\alpha_i)$ and are subdominant. From this perspective, it might be more appropriate to refer to \eqref{eq:massentropyrelation} as an area/extremality relation.

\section{Energy Conditions and Extremality Corrections} \label{sec:ExtremalCorrections}
The fact that the mass correction is given by the integral over the effective stress suggests a possible relation between the WGC and energy conditions on the perturbations we are considering, at least in cases where the notion of the effective stress tensor coincides with the usual stress tensor. As mentioned before, this happens when
\beq
\frac12g_{ab}F_{cd}\delta F^{cd} - F_{ac}\delta F_b^{\,\,c} = 0 ~.
\eeq
Notably, this is the case for Kerr black holes, but after performing a contraction with $K^a$ we find that this term also vanishes for electrically charged Reissner-Nordström black holes. Thus, in these cases we find that the correction to the mass in a canonical ensemble is given by
\beq
\lim_{T\to0}\left(\delta M\right)_{T,J,Q_e} = \lim_{T\to0}\int_\Sigma\tilde\epsilon_{bcd}\delta T_{ae}K^an^e ~,
\eeq
A decrease in the mass (as required for the WGC to be satisfied by charged higher-derivative corrected black holes) therefore implies that at least at some point along the Cauchy surface $\delta T_{ab}K^an^b <0 $. When $K^a$ and $n^a$ are two cooriented timelike vectors, this corresponds to a violation of the DEC for the stress tensor of the perturbation, which is $\delta T_{ab}$. Because $n^a$ is the unit normal to the Cauchy surface it is a future pointing timelike vector by definition. Similarly, away from the horizon the Killing vector $K^a$ is everywhere future-pointing and timelike for the Reissner-Nordström black hole, but not for Kerr black holes. In that case, the nature of $K^a$ depends on the azimuthal angle $\theta$. However, for an energy condition to be violated it is sufficient to find a violation at just a single point so for simplicity we can focus on the surface $\theta=0$. In that case, $K^a$ is everywhere timelike away from the horizon and we come to the same conclusion that the matter generating $\delta T_{ab}$ must violate the DEC to satisfy the WGC. Here, we should make it clear that a violation of the DEC is only required for the additional matter perturbing the (uncorrected) black hole. Typically, the total stress tensor which includes the (unperturbed) electromagnetic stress tensor does satisfy the DEC, simply because of the positive energy density of the electric field which is only perturbed slightly by adding matter.

However, while a violation of the DEC is a necessary condition for the WGC to be satisfied it is not sufficient. We still need to perform the integral over the Cauchy surface and one could imagine a situation where $\delta T_{ab}K^an^b < 0 $ at some point(s), but the entire integral gets overwhelmed by positive contributions in such a way to yield a positive result. Prohibiting this situation either requires the contraction of the stress tensor to have a definite sign or a more general energy condition that bounds the integral over the stress tensor. However, integrated energy conditions typically hold for timelike or null trajectories \cite{Kontou:2020bta}, so it seems difficult to bound the integral over the spacelike Cauchy surface this way. We will now examine the stress tensor and extremality corrections corresponding of higher-derivative corrected Reissner-Nordström and Kerr black holes.

\subsection{AdS-Reissner-Nordström Black Hole}
We now consider an electrically charged Reissner-Nordström black hole in AdS space perturbed by the following parity-even four-derivative higher-derivative operators.
\beq
\bal
L &= \Big(\frac12R+\frac{3}{\ell^2}-\frac14F_{ab}F^{ab} + \frac{a_1}{4}(F_{ab}F^{ab})^2 \\
&\quad+ \frac{a_2}{2}F_{ab}F_{cd}W^{abcd} \Big)\epsilon~.
\eal
\eeq
We note that this is not the most general set of leading higher-derivative corrections, which would require us to include a Weyl-squared operator $W_{abcd}W^{abcd}$. However, in AdS space this term falls off as ${\cal O}(1/r^4)$ at large $r$, which is not fast enough to not contribute to a boundary term that was an assumption in the derivation of \eqref{eq:Iyer-Waldrelation}. Therefore additional care must taken to take into account the Weyl-squared term and for simplicity we will drop it and leave this possibility for future work. In \cite{Cremonini:2019wdk} a thermodynamic approach based on holographic renormalization was used to compute the effect of such corrections. Nonetheless, if we consider black holes that are small with respect to the AdS radius, it is justified to ignore the Weyl-squared term because in that case it can be rewritten in terms of the other operators using the equations of motion.

The metric and gauge field that solve the equations of motion at the two-derivative level are given by
\beq
\bal
\rmd s^2 &= -f(r)\rmd t^2 + f(r)^{-1}\rmd r^2 + r^2\rmd \Omega_{k,2}^2 \\
f(r) &= k - \frac{M}{4\pi} + \frac{Q^2}{32\pi^2r^2} + \frac{r^2}{\ell^2} \\
A &= -\frac{Q}{4\pi r}\rmd t ~.
\eal
\eeq
The parameter $k=(-1,0,1)$ determines the horizon geometry that is, respectively, hyperbolic, planar or spherical. Thus, for $k=0$ this metric describes a black brane. $\rmd\Omega_{k,2}^2$ denotes the respective area element. As discussed, we take a gauge for the gauge field in which it vanishes at infinity. It will be useful to write $f(r)$ in terms of its roots as
\beq
\bal
f(r) &= \frac{(r+r_0) (r-r_-) (r-r_+) (r-r_0+r_-+r_+)}{\ell^2 r^2} ~,\\
2r_0 &=\sqrt{-4 k \ell^2-3 r_-^2-2 r_-r_+-3 r_+^2}+r_-+r_+ ~.
\eal
\eeq
The relationship with the ADM mass and electric charge is
\beq
\bal
Q^2 &= \frac{32 \pi ^2 r_0r_- r_+ (r_-+r_+-r_0)}{\ell^2} ~, \\
M &= \frac{4 \pi(r_+-r_0)(r_0-r_-) (r_-+r_+)}{\ell^2}.
\eal 
\eeq
Varying with respect to the metric and gauge field, we obtain the expression for the perturbed stress tensor
\beq
\bal
&\delta T^g_{ab} = \frac{a_1}{4}\left(g_{ab}(F_{cd}F^{cd})^2-8F_{cd}F^{cd}F_a^{\,\,\,e}F_{be}\right) \\
&+ \frac{a_2}{2}\bigg[g_{ab}R_{cdef}F^{cd}F^{ef}-6F_{cb}F^{de}R^c_{\,\,\,ade} \\
&-4\nabla_d\nabla_c\big(F^c_{\,\,\,a}F^d_{\,\,\,b}\big) -2g_{ab}R^{cd}F_{ce}F_{d}^{\,\,\,e} +8R_{bc}F_{ad}F^{cd} \\
&+ 4 R^{cd}F_{ca}F_{db} +2 g_{ab} \nabla_c\nabla_d\big(F^c_{\,\,\,e}F^{de}\big) -4\nabla_c\nabla_b\big(F_{ad}F^{cd}\big) \\
&+2\square\big(F_{ac}F_{b}^{\,\,\,c}\big)+\frac13g_{ab}RF_{cd}F^{cd} -\frac43RF_{a}^{\,\,\,c}F_{bc} \\
&-\frac23F_{cd}F^{cd}R_{ab}+\frac23\nabla_a\nabla_b\big(F_{cd}F^{cd}\big) -\frac23g_{ab}\square\big(F_{cd}F^{cd}\big) \bigg] ~,
\eal
\eeq
and the electric source
\beq
j^a = 2\nabla_b\left(a_1F_{cd}F^{cd}F^{ab} + a_2W^{abcd}F_{cd} \right) ~.
\eeq
To check if a violation of the DEC is a sufficient condition for the WGC to be satisfied, we can evaluate $\delta T_{ab}n^aK^b$. In general, this leads to a rather lengthy expression, so we summarize the situation. First of all, we note that for $k=-1$ the stress tensor always has an indefinite sign for fixed Wilson coefficients due to the negative curvature of the horizon. Also, for $k=(0,1)$ the term proportional to $a_2$ similarly does not have a definite sign, but the term proportional to $a_1$ is given by
\beq
\bal
&\left.\frac{5\ell^4}{12r_+^3\Omega_{0,2}}\delta T_{ab}n^aK^b\right|_{a_2=0} = -\frac{15r_+^5\sqrt{r^4-4 r r_+^3+3r_+^4}}{\ell r^9}a_1 ~, \\
&\left.\frac{5\ell^4}{12r_+^3\Omega_{1,2}}\delta T_{ab}n^aK^b\right|_{a_2=0} = -\frac{5 r_+\left(\ell^2+3 r_+^2\right)^2 (r-r_+) }{3 \ell r^9}a_1 \\
&\qquad\qquad\qquad\qquad\qquad\quad\times\sqrt{\ell^2+r^2+2 r r_++3 r_+^2} ~.
\eal
\eeq
$\Omega_{k,2}$ denotes the unit area element, which diverges for $k=(-1,0)$. In these cases, one can still consider corrections to the energy and entropy density. Thus, a violation of the DEC only fixes the sign of the contribution of the $(F_{ab}F^{ab})^2$ operator. We note that the condition $a_1 \geq 0$ also follows from unitarity, while $a_2$ cannot be argued to take a particular sign this way \cite{Hamada:2018dde,Arkani-Hamed:2021ajd}. However, in addition to the DEC we can also consider the NEC. Using the null vector $N^a = (\sqrt{-1/g_{tt}},\sqrt{1/g_{rr}},0,0)$, we find
\beq
k=(0,1):\qquad \delta T_{ab}N^aN^b \geq 0 \quad \to \quad a_2 \leq 0 ~,
\eeq
To see how these conditions can be used to constrain the corrections to the extremality bound we now perform the integral over the stress tensor.
\beq
\bal
&\frac{5\ell^4}{12r_+^3\Omega_{k,2}}\int_\Sigma\tilde\epsilon_{bcd}\delta T_{ae}K^an^e = \\
& -\frac{\left(k \ell^2+3 r_+^2\right) \left(2 a_1 \left(k \ell^2+3 r_+^2\right)-a_2 \left(k \ell^2-2 r_+^2\right)\right)}{6 r_+^4} ~.
\eal
\eeq
For $k\neq 0$ the combination of Wilson coefficients that determines the correction differs for small and large black holes.\footnote{Here, small or large refers to the size of the black hole with respect to the AdS radius. The small black holes we are studying should not be confused with black holes whose horizon size is so small that the higher-derivative expansion breaks down.} Small black holes ($r_+/\ell \ll 1$) with either a spherical or hyperbolic horizon ($k=\pm1$) don’t notice the AdS curvature, so the combination of Wilson coefficients controlling corrections to the extremality bound is the same as in flat space \cite{Kats:2006xp}
\beq
\frac{5\ell^4}{12r_+^3\Omega_{\pm1,2}} \int_\Sigma\tilde\epsilon_{bcd}\delta T_{ae}K^an^e =-\frac{\left(2 a_1-a_2\right) \ell^4}{6 r_+^4} ~.
\eeq
On the other hand, for large black holes $r_+/\ell \gg 1$ and we find that
\beq
\frac{5\ell^4}{12r_+^3\Omega_{\pm1,2}} \int_\Sigma\tilde\epsilon_{bcd}\delta T_{ae}K^an^e = -\left(3a_1+a_2\right) ~,
\eeq
which is also the correction for black branes:
\beq
\frac{5\ell^4}{12r_+^3\Omega_{0,2}} \int_\Sigma\tilde\epsilon_{bcd}\delta T_{ae}K^an^e = -\left(3a_1+a_2\right) ~.
\eeq
Note that for hyperbolic black holes ($k=-1$) there is a particular size for which all corrections vanish, that is $r_+^2=\ell^3/3$, and for spherical black holes ($k=+1$) the correction proportional to $a_2$ vanishes when $r_+^2=\ell^2/2$. The sign of the corrections to the mass in a canonical ensemble is therefore controlled by a different combination of Wilson coefficient depending on the horizon geometry and the size of the black hole.

For small black holes in AdS space for which ignoring the Weyl squared term is justified we find that unitarity $(a_1\geq 0)$ in combination with the NEC $(a_2\leq 0)$ requires the extremal mass to decrease in a canonical ensemble, consistent with the WGC.

\subsection{Kerr Black Hole}
We also consider the leading higher-derivative corrections to the Kerr black hole. Although there is no WGC-like motivation that constrains the sign of higher-derivative corrections to Kerr black holes, due to the Penrose process extremal Kerr black holes are unstable, it is still interesting to see how their properties are corrected. The leading parity-even higher-derivative term has six derivatives and the Lagrangian is given by\footnote{Additional terms contribute if one includes light degrees of freedom in addition to the metric \cite{Cano:2019ore}.}
\beq
L=\frac1{2}\left(R+\alpha R_{abcd}R^{cdef}R_{ef}^{\,\,\,\,\,ab}\right)\epsilon ~.
\eeq
At the two-derivative level, the Kerr solution is given by \eqref{eq:KerrNewman} with $r_Q=0$. Performing the variation with respect to the metric, we find that the perturbed stress tensor is
\beq
\bal
\frac{\delta T_{ab}}{\alpha}&= \frac12g_{ab}W_{cd}^{\,\,gh}W^{cdef}W_{efgh}- 6(\nabla_cW_{bfde})(\nabla^fW_a^{\,\,cde}) \\
&+ R_{a}^{\,\,cde}\left[3R_{bd}^{\,\,fg}R_{cefg} + 12{R_b^{\,\,f}}_{d}^{\,\,g}R_{cfeg}-3R_{bc}^{\,\,fg}R_{defg} \right. \\
&\left.+3{R_b^{\,\,f}}_{c}^{\,\,g}R_{defg}\right]   ~.
\eal
\eeq
The corrections are now easily found by evaluating the stress tensor. As we already remarked before, due to the dependence on the azimuthal angle $\theta$ the contraction of the stress tensor $\delta T_{ab}n^aK^b$ does not have a definite sign for a fixed $\alpha$. This means that violation of the DEC is a necessary but not sufficient condition to fix the sign of the mass correction in a canonical ensemble. For the same reason, the NEC is always violated for some value of $\theta$.

Performing the integral we find
\beq
\bal
&\int_\Sigma\tilde\epsilon_{bcd}\delta T_{ae}K^an^e  = \\
& \frac{4 \pi  \alpha}{35 r_+^3 (r_-+r_+)^3}\left(-37 r_-^3+357 r_-^2 r_+-455 r_- r_+^2+175 r_+^3\right) ~.
\eal
\eeq
Interestingly, for fixed $\alpha$ the right-hand side does have a definite sign which implies that the correction to the Bekenstein-Hawking entropy also has a definite sign even away from extremality. In the extremal limit the correction reduces to
\beq
\int_\Sigma\tilde\epsilon_{bcd}\delta T_{ae}K^an^e = \frac{4 \pi \alpha }{7 r_+^3} ~.
\eeq
Reproducing the result in \cite{Reall:2019sah}. We see that at extremality the mass in a canonical ensemble decreases when $\alpha<0$. However, in contrast to charged black holes no energy condition fixes the sign of $\alpha$ to be negative. Perhaps, this is not so surprising because, as discussed, there is no WGC-like motivation that constrains the sign of higher-derivative corrections to Kerr. Only when the rotation of a black hole can be related to charge, such a connection might arise \cite{Aalsma:2019ryi,Aalsma:2020duv}.

\section{Weak Gravity Conjecture in Anti-de Sitter Space} \label{sec:AdSWGC}
We will now consider the derived corrections to the extremality bound of AdS black holes in the light of a formulation of the WGC in AdS space. For (electrically) charged asymptotically flat black holes it is clear that the mild form of the WGC requires higher-derivative corrections to decrease the mass in a canonical ensemble. This has been used to derive bounds on Wilson coefficients. Vice versa, the WGC has been derived by using bounds on Wilson coefficients coming from unitarity and causality. 

However, for black holes that are asymptotically AdS the correct formulation of the WGC is less clear although proposals exist, such as \cite{Harlow:2015lma,Montero:2018fns}. Recently, a version of the WGC in AdS space has been formulated \cite{Aharony:2021mpc} that is inspired by the repulsive force conjecture (RFC) \cite{Palti:2017elp,Heidenreich:2019zkl}. The RFC states that, for two identical charged particles, their total long-range force should be repulsive rather than attractive (or vanish). For particles that only experience gravitational and electromagnetic forces, this conjecture is identical to the WGC in flat space. However, when massless scalar fields are included the RFC and WGC become distinct statements. It is now interesting to understand if there is a mild version of the RFC where the self-repulsive states are black holes themselves. In flat space, this question has recently been discussed in \cite{Cremonini:2021upd}, where it was shown in several examples that the RFC cannot be satisfied by black holes along all charge directions. This calls into question the validity of a mild form of the RFC.

\begin{figure}[h]
\centering
\includegraphics[scale=1.7]{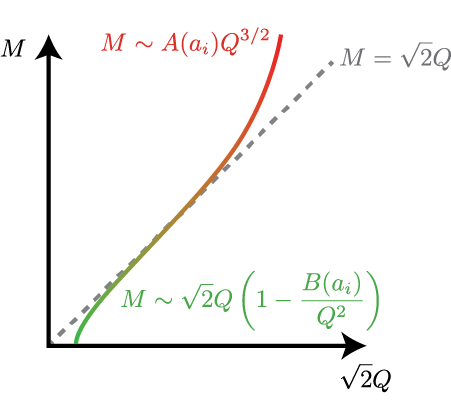}
\caption{Cartoon of the extremality bound of spherical black holes in AdS. Higher-derivative corrections lead to a modification of the extremality bound. The mass of large extremal black holes in AdS (red) is a convex function of charge irrespective of the sign of the Wilson coefficients $a_i$. The mass of small black holes in AdS (green) is a concave function of charge if they satisfy the WGC ($B(a_i)>0$).}
\label{fig:LargeSmallBHs}
\end{figure}

Also, in AdS space formulating the RFC is difficult because there is always a contribution from the cosmological constant to the force acting on a state. Therefore, \cite{Aharony:2021mpc} (see also \cite{Antipin:2021rsh}) suggested a positive binding conjecture, which requires the existence of a charged state that has non-negative self-binding energy to avoid the formation of a large tower of bound states. In the dual conformal field theory (CFT) this was phrased as a convexity condition for the scaling dimensions of charged CFT operators. Given this proposal, one might wonder if it possible that a mild form of this convex charge conjecture (CCC) is satisfied by higher-derivative corrected black holes in AdS. Although the CCC does not necessarily require black hole states to satisfy the convexity condition, it is an interesting questions to ask if they could. In fact, it was already observed in \cite{Aharony:2021mpc} that large black holes in AdS, dual to heavy operators in the CFT, are convex. So, does convexity put any constraints on the sign of the Wilson coefficients of higher-derivative operators? For large black holes in AdS$_4$ ($r_+/\ell \gg 1$) the extremality bound scales as $M\sim A(a_i) Q^{3/2}$, where $A(a_i)$ is a charge-independent positive constant that depends on the Wilson coefficients, see Fig. \ref{fig:LargeSmallBHs}. 
On the other hand, small extremal black holes in AdS space are marginally convex at the two-derivative level since their mass scales linearly with charge. Now, higher-derivative corrections that are consistent with the WGC decrease the mass ($B(a_i)>0$) at fixed charge resulting in a concave function. As observed in \cite{Aharony:2021mpc} this implies that, when the WGC is satisfied for small black holes, they violate the convexity condition. However, since these black holes are unstable they do not lead to a large tower of stable states. On the other hand, large black holes are always convex irrespective of the sign of the Wilson coefficients. Therefore, although the CCC is automatically satisfied for large black holes in AdS, it does not put any constraints on the sign of Wilson coefficients of higher-derivative corrections.\\

\section{Conclusions} \label{sec:conclusion}
In this paper, we considered corrections to extremal black holes by arbitrary perturbations using the Iyer-Wald formalism. The main result is a relation between the correction to the extremality bound and an effective stress tensor. In particular cases of interest, which we clarified, the effective stress tensor reduces to the usual stress tensor. Furthermore, we gave a new derivation of the entropy/extremality relation that has appeared before in the literature. Because we relate corrections to the extremality bound to the stress tensor, it clarifies the connection between the WGC and energy conditions. In particular, we showed that a necessary condition to decrease the mass of an extremal black hole in a canonical ensemble is that the perturbation violates the DEC. Combining a unitarity bound with the NEC we showed that the mild form of the WGC follows for charged black holes.

In future work, it would be interesting to generalize the relation we derived to include black holes with magnetic charges and couplings to additional (scalar) fields. This requires some special attention because those contributions do not appear in the standard way the Iyer-Wald formalism is applied \cite{Mitsios:2021zrn}. Additionally, to derive our main result we assumed that the perturbations we considered decay sufficiently fast at infinity. Notably, this is not the case for all four-derivative operators in AdS space and taking into account these perturbations, which will contribute additional boundary terms, also seems like a worthwhile endeavor. Moreover, it would be interesting to understand better the relation between energy conditions and the WGC. In particular, while it is reasonable to expect that higher-derivative corrections generated by integrating out healthy matter at tree level satisfies (some) energy conditions, quantum effects are known to violate them. Nonetheless, in that case it is still expected that some averaged condition holds \cite{Kontou:2020bta}.

To conclude, we hope that the general method we presented here to compute extremal black hole corrections will provide useful in further studies of black holes and the WGC.

\section*{Acknowledgments}

I am grateful to Alex Cole, Eleni Kontou, Greg Loges and Gary Shiu for useful discussions. I especially thank Greg Loges for carefully explaining some of the subtleties of black hole thermodynamics and Eleni Kontou for feedback on an earlier version of this paper. This work is supported by the DOE under Grant No. DE-SC0017647.

\appendix

\bibliographystyle{apsrev4-2}
\bibliography{refs}

\end{document}